\documentclass{article}
\usepackage{PRIMEarxiv}

\usepackage[utf8]{inputenc} 
\usepackage[T1]{fontenc}    
\usepackage[hidelinks]{hyperref}       
\usepackage{natbib}
\usepackage{bibentry}
\usepackage{appendix}
\usepackage[dvipsnames]{xcolor} 
\usepackage{url}            
\usepackage{booktabs}       
\usepackage{amsmath, amsthm, amssymb, amsfonts, mathtools, mathrsfs,bm, xfrac}       
\usepackage{nicefrac}       
\usepackage{lipsum}
\usepackage{microtype}      
\usepackage{fancyhdr}       
\usepackage{graphicx}       
\graphicspath{{media/}}     
\numberwithin{equation}{section} 

\newcommand{\mc}[1]{\mathcal{#1}} 
\newcommand{\p}[0]{\partial} 
\newcommand{\Real}[0]{\text{Real}} 
\newcommand{\Imag}[0]{\text{Imag}} 

\title{Multi-domain AAA-least squares methods and applications}

\author{
  Samuel J. Harris, Ryan A. Palmer \\
  School of Engineering Mathematics and Technology \\
  University of Bristol \\
  University Walk, Bristol, BS8 1TW, UK\\
  \texttt{samuel.harris@bristol.ac.uk, ryan.palmer@bristol.ac.uk}}

\begin{document}
\maketitle

\begin{abstract}
A multi-domain AAA-least squares (nAAALS) method is developed for computing rational approximations of harmonic functions defined in the interior and exterior of multiply-connected domains. Three classes of problems are considered: i) multi-domain problems where an unbounded exterior is punctured by multiple singly-connected domains; ii) nested domain problems where one domain contains multiple punctured domains; and iii) quasi-steady time-stepping problems where the nAAALS algorithm is coupled with standard Runge-Kutta schemes to compute the coupled evolution of the multiple domain boundaries. The nAAALS method extends the original AAALS algorithm in two key ways. First, a function $\tilde{S}$ satisfying $\tilde{S}(z)=\overline{z}$ on each domain boundary is approximated using the AAA algorithm, producing simple poles distributed according to both the individual boundary geometries and the global multi-domain configuration. Second, the least-squares framework is modified to accommodate the coupled matching conditions across each domain boundary. The method is applied to problems in vortex dynamics to demonstrate its capability, focusing specifically on vortex patches as a working example: finite-area regions of incompressible, inviscid fluid in rotational flow. The nAAALS algorithm is used to (a) reproduce known vortex patch equilibria, and (b) compute the evolution of new and existing non-equilibrium configurations. The results are shown to be accurate at low computational cost, running in seconds on a standard laptop. Crucially, the  nAAALS method is suitable for addressing a broader class of steady and quasi-steady multi-domain problems across disciplines, such as those arising in applied mathematics, mathematical biology, and fluid dynamics.
\end{abstract}


\section{Introduction}\label{sec:introduction}
The past eight years have seen the development of various numerical methods founded upon the principle of rational approximation in two dimensions. Approximating a function $f(z)$ by a rational function $r(z)=p(z)/q(z)$, where $p$ and $q$ are polynomials, can provide greater accuracy than approximating by a polynomial alone - the special case where $q(z)\equiv1$. Rational approximation offers significant advantages when the target function contains singularities on or near the approximation domain, a feature that is difficult to capture with polynomial approximation. Three distinct, yet strongly related, numerical methods, which utilise rational approximation, have recently emerged - a comprehensive review of these three methods is given \cite{nakatsukasa2026applications} but they are briefly summarised below.

Lightning solvers provide rational approximations of functions satisfying the Laplace \citep{gopal2019solving}, Helmholtz \citep{gopal2019new} or biharmonic \citep{xue2024computation} equations in simply-connected domains. The key idea is that these algorithms manually cluster simple poles exponentially close to pre-defined corners and cusps in the domain boundary and use them in the rational approximation, matching to boundary data via a least-squares fitting. The AAA algorithm (adaptive Antoulas-Anderson, pronounced `triple A') introduced in \cite{nakatsukasa2018aaa} calculates a rational approximation $r$ of a function $f$ on a given real or complex domain $Z$. The approximation $r$ is represented in barycentric form, with its barycentric weights computed by linear least-squares and its support points selected through a systematic greedy procedure. While this approximation is highly accurate (close to machine precision) on the domain $Z$ itself, its accuracy deteriorates away from $Z$.

The focus of this paper is the third method: the AAA-least squares (AAALS) algorithm which is a combination of the previous two methods by incorporating the AAA algorithm into lightning solvers. Rather than placing poles manually, the AAA algorithm is used to generate suitable pole locations for a given domain boundary, which are then used as inputs into the lightning solver and the least-squares fitting proceeds as usual. This approach circumvents the need to specify corners and cusps in the boundary a priori and produces a suitable set of poles in a single run of the AAALS method. As the AAA algorithm produces poles both interior and exterior to the boundary, the only additional step is to discard any poles lying within your approximation domain to avoid introducing singular behaviour into the rational approximation. 

Originally, the AAALS method was developed to solve problems posed on interior or exterior singly- and multiply-connected domains, see \cite{costa2023aaa}. The algorithm has since been extended to two-domain problems, where the interior and exterior domains are both considered with homogeneous \citep{costa2024modelling} or inhomogeneous \citep{harris2025modeling} matching boundary conditions. This has allowed the AAALS method to be applied to a range of interdisciplinary problems in mathematical biology \citep{harris2023penguin}, fluid dynamics including potential, Hele-Shaw and Stokes flows \citep{xue2024computation,baddoo2025computing}, electro- and magnetostatics \citep{costa2024modelling,harris2025modeling}, and wildfire modelling \citep{harris2025modelling}.

In this work, we present a multi-domain AAALS algorithm capable of computing rational approximations of harmonic functions both interior and exterior to various multiply-connected configurations consisting of $N>0$ domains, hence we name this method the `nAAALS' algorithm. Fundamentally, this extension unifies the previously developed multiply-connected and two-domain extensions to AAALS method, using the same core framework as the original one-domain algorithm \cite{costa2023aaa} that is then adapted to account for the additional domains (and thus boundaries). As the nAAALS method remains computationally efficient, this motivates us to also couple the algorithm to standard Runge-Kutta (RK) time-stepping methods \citep{dormand1980family}. This allows the nAAALS method to then be used to solve quasi-steady, time-varying problems, for example free boundary problems in fluid dynamics \citep{bazant2005conformal}, Stefan problems involving melting, freezing or dissolving interfaces \citep{rycroft2016asymmetric, ladd2020}, and the tracking of an evolving wildfire front \citep{hilton2018incorporating,harris2025modelling}. 

While much of the original AAALS framework from \cite{costa2023aaa} is retained in this work, we follow \cite{costa2024modelling} in making a significant change in how poles are generated. The original one-domain AAALS method in \cite{costa2023aaa} generates poles by using the AAA algorithm to approximate the Dirichlet boundary condition itself. Similarly, the two-domain AAALS method in \cite{harris2025modeling} uses the AAA algorithm twice to approximate the two functions representing the Dirichlet and Neumann boundary conditions and then combining both sets of poles. While there are potential occurrences of pole `overcounting' and `undercounting', these cases are inconsequential as any exponential clustering of poles near corner singularities is suitable to use in the AAALS method \citep{gopal2019solving}. When considering the one- and two-domain AAALS methods in our current work, we instead generate poles by using the AAA algorithm to approximate the Schwarz function $S(z)$, the complex analytic function satisfying $S(z)=\overline{z}$ on the domain boundary. Geometric features of the boundary curve are encoded into the Schwarz function itself, allowing poles associated with corner and cusp singularities to be identified using the boundary data alone, independent of the particular boundary conditions of the Laplace problem. This approach also uses the AAA algorithm only once in both the one- and two-domain AAALS methods, amounting to a (small) decrease in computation time for the two-domain method. 

For the nAAALS method, poles are also generated using only one run of the AAA algorithm to approximate the function $\tilde{S}(z)$ satisfying $\tilde{S}(z)=\overline{z}$ on all $N$ boundaries. To the author's knowledge, such a function has not been defined or used previously in the literature, though its boundary conditions suggest it contains geometrical information about multiply-connected domains in a similar way to the Schwarz function for singly-connected domains. While this new function is not explored in-depth in this work, we use it to generate poles which are adapted both to the individual boundary geometries and the overall multi-domain configuration. The only notable complication in using the functions $S(z)$ and $\tilde{S}(z)$ to generate poles arises if the boundary conditions contain singularities that do not correspond with singularities in the boundary geometry. Such situations are expected to be rare (indeed, none occur in the examples considered in our current work) and would typically be known a priori. In these cases, additional AAA or lightning poles could be supplied by the user near these corresponding regions. 

To validate the effectiveness of the proposed nAAALS algorithm, we apply the method to a range of problems from vortex dynamics, in a similar approach to \cite{harris2025vortex}. In particular, we consider vortex patches, finite-area regions of incompressible, inviscid fluid in rotational flow. Configurations of vortex patches that steadily translate or rotate over time are known as vortex equilibria, of which various analytical and numerical equilibrium solutions are known in the literature, see e.g. \citep{pierrehumbert1980family,wu1984steady,morel1994multipolar,xue2017new}. The nAAALS algorithm can be used in conjunction with iterative schemes developed by \cite{wu1984steady} and \cite{xue2017new} to reproduce these known vortex patch equilibria. Furthermore, the nAAALS + time-stepping (nAAALS + TS) framework can be used to compute the evolution of non-equilibrium vortex patch configurations. Crucially, the aim of this work is not to discover new analytical or numerical vortex patch equilibria, nor to study in detail the evolution of non-equilibrium configurations. Instead, existing results from vortex dynamics are here reproduced to demonstrate the efficacy of the new nAAALS methodology developed in this work.

This paper is organised as follows. Section \ref{sec:onetwodom} summarises the existing one- and two-domain AAALS algorithms. Section \ref{sec:NdomAAALS} introduces the nAAALS algorithm and uses it to reproduce known translating and rotating vortex patch equilibria in section \ref{sec:multidom}. Section \ref{sec:nested} develops a variant of the nAAALS method for nested domains, a bounded domain punctured by one or more interior domains, which is then used to compute the velocity field of equilibrium and non-equilibrium nested vortex patch configurations. The nAAALS method is then coupled to RK time-stepping methods in section \ref{sec:time-stepping} and section \ref{sec:discussion} summarises the new multi-domain extensions to the AAALS method, outlining possible directions for future research.

\section{One- and two-domain AAALS methods}\label{sec:onetwodom}
We first summarise the standard one- and two-domain versions of the AAALS method. In a one-domain context, consider the following Laplace problem for a harmonic function $\psi_1$ in some two-dimensional domain $\Omega_1$ bounded by the simple, closed curve $\gamma_1$,
\begin{equation}\label{eq:onedompb}
    \nabla^2\psi_1 = 0\;\text{in } \Omega_1, \quad\quad \psi_1 = h(z)\;\text{on }\gamma_1, 
\end{equation}
where $h$ is some given function. Following the work of \cite{costa2023aaa}, the harmonic function can be written as $\psi_1=\Real[F_1(z)]$, where the complex analytic function $F_1(z)$ has the following rational approximation
\begin{equation}\label{eq:onedomrat}
    F_1(z) \approx \sum_{m=0}^Ma_m(z-C_1)^m+\sum_{m=1}^{M_1}\frac{b_m}{z-p_m}.
\end{equation}
The first (Runge) term is a smooth, truncated power series, where $M$ is the series truncation and $C_1$ is some point in $\Omega_1$ typically taken to be the centre. The second (Newman) term consists of $M_1$ simple poles $p_m$ exterior to $\Omega_1$ distributed exponentially close to corners and cusps in $\gamma_1$ to represent the singular behaviour occuring at these points. To find suitable poles for a given domain geometry, the AAA algorithm is used to produce a rational approximation of the Schwarz function $S(Z)=\overline{Z}$ for given boundary data $Z$ \citep{trefethen2025numerical}. Finally, to find the unknown complex coefficients $(a_m, b_m)$, the boundary condition in \eqref{eq:onedompb} is used to construct the matrix product $Ac=H$, where $A$ is a matrix of basis vectors of the boundary data and the solution vector $H$ is given by $H=h(Z)$. The vector of unknown coefficients $c$ can then be found by a least-squares fitting, for example using the MATLAB backslash operation $c=H\backslash A$. Note that this one-domain AAALS method can also be used when the Dirichlet boundary condition $\psi_1=h(z)$ is replaced by the Neumann boundary condition $\psi_{\hat{n}_1}=0$, where the subscript $\hat{n}_1$ denotes differentiation with respect to the outward unit normal vector $\hat{n}_1$ (in complex notation) of $\gamma_1$, and when considering Laplace problems in exterior domains $\Omega_0=\mathbb{C}\backslash\Omega_1$ by instead using negative powers in the Runge term.

For a two-domain scenario, consider again the two-dimensional bounded domain $\Omega_1$ enclosed by the curve $\gamma_1$ with outward unit normal vector $\hat{n}_1$. The function $\psi$ is defined exterior ($\psi=\psi_0$) and interior ($\psi=\psi_1$) to $\gamma_1$, satisfying Laplace's equation in the exterior region $\Omega_0=\mathbb{C}\backslash\Omega_1$ and a Poisson equation with constant right hand side $\omega$ in the interior region $\Omega_1$. To non-dimensionalise the system, we introduce the scalings $z=Lz^\ast$ and $\psi = (\omega L^2)\psi^\ast$, where starred quantities are dimensionless and $L$ is the (conformal) radius of $\Omega_1$. Dropping stars immediately and applying suitable matching Dirichlet and Neumann boundary conditions and the far-field condition $\psi_0\rightarrow \tilde{G}_0(z)$ for some function $\tilde{G}_0$ yields the following system
\begin{gather}
    \nabla^2\psi_0=0\;\;\;\text{ in }\Omega_0,\label{Meq:eGE}\\
    \nabla^2\psi_1=1\;\;\;\text{ in }\Omega_1,\label{Meq:iGE}\\
    \psi_0=\psi_1\;\;\;\text{ on $\gamma_1$,}\label{Meq:dBC}\\
    \frac{\p \psi_0}{\p \hat{n}_1} = \mu_1\frac{\p\psi_1}{\p\hat{n}_1}\;\;\;\text{ on $\gamma_1$,}\label{Meq:nBC}\\
    \psi_0\rightarrow \tilde{G}_0(z) \;\;\;\text{as $r\rightarrow\infty$,}\label{Meq:far}
\end{gather}
where $\mu_1$ is a constant.

Rational approximations for both the interior and exterior stream functions can be formulated, in particular $\psi_0 = \Real[G_0(z)] + \Real[F_0(z)]$ and $\psi_1=|z-C_1|^2/4 + \Real[F_1(z)]$, where $C_1$ is the dimensionless centre of $\Omega_1$. The functions $F_n(z)$ are of the form \eqref{eq:onedomrat} with positive/negative exponents used in the power series for the interior/exterior function, respectively. The additional terms account for the far-field condition \eqref{Meq:far} and the non-zero right hand side in \eqref{Meq:iGE}, noting that $\tilde{G}_0(z)=\Real[G_0(z)]$. Once again, poles are found by approximating the Schwarz function $S(z)$ by the AAA algorithm, with interior poles used in $F_0(z)$ and exterior poles in $F_1(z)$. The unknown complex coefficients can then be found by solving the coupled boundary conditions \eqref{Meq:dBC} and \eqref{Meq:nBC}, which can be rewritten in terms of $F_0(z)$ and $F_1(z)$ as
\begin{gather}
    \Real[F_0(z)]-\Real[F_1(z)] = -\Real[G_0(z)]+\frac{|z-C_1|^2}{4},\label{eq:twodomdBC}\\
    \Real[\hat{n}_1F_0'(z)]-\mu_1\Real[\hat{n}_1F_1'(z)] = -\Real[\hat{n}_1G_0'(z)]+\frac{\mu_1}{2}\Real[\hat{n}_1\overline{(z-C_1)}].\label{eq:twodomnBC}
\end{gather}
noting that $\p\psi/\p\hat{n}_1 = \Real[\hat{n}_1\overline{\nabla}\psi]$ and $\overline{\nabla}[\Real[F(z)]] = F'(z)$ as in \cite{trefethen2018series}. These two boundary conditions can then be represented as a matrix product $Ac=H$, where $H$ includes the right-hand sides of both \eqref{eq:twodomdBC} and \eqref{eq:twodomnBC}, $A$ is the matrix of basis vectors for the left-hand sides, and $c$ are the unknown coefficients for both $F_0$ and $F_1$. Once more, the backslash operation is used to solve $c=H\backslash A$. 

Fundamentally, the two-domain AAALS method utilises the same framework as its one-domain analogue but with two key differences. First, all of the poles found by the AAA algorithm are now used, taking care to use interior poles in the rational approximation of the exterior function and vice versa. Second, as we have two boundary conditions, the matrix product $Ac=H$ is approximately twice as large in both dimensions. Suppose each boundary consists of $P$ points, then the dimensions of matrix $A$ are $2P\times M_T$ where $M_T=2[2(M+1)+M_0+M_1]$ and $M_{0|1}$ are the number of poles exterior to region $\Omega_{0|1}$, respectively\footnote{The additional factor of two present in the equation for $M_T$ is because we are solving for complex coefficients.}. Practically, the computational runtime of the two-domain AAALS algorithm is thus approximately twice that of the one-domain method, meaning it remains computationally efficient (on the order of seconds) on a standard laptop.

Before proceeding, several important factors of the numerical method are worth highlighting. First, while it is possible to consider only the singular Newman part in the rational approximation \eqref{eq:onedomrat}, it was shown in \cite{herremans2023resolution} that including a smooth Runge term improves the numerical accuracy of the rational approximation. In their work, the series truncation $M=\lceil1.3\sqrt{M_0}\rceil$ is used; in our current work, typically $\mathcal{O}(100)$ poles are produced for all geometries considered and hence a representative series truncation of $M=20$ is used in all upcoming sections. Second, the Runge part of the Vandermonde matrix $A$ is generally ill-conditioned, and hence it was proposed in \cite{brubeck2021vandermonde} that instead of orthogonalising $A$ in a QR factorisation, the matrix can be sequentially orthogonalised during each step of its construction in the standard (Gram-Schmidt) manner. The resulting matrix in this so-called Vandermonde with Arnoldi orthogonalisation process is well-conditioned, even in complicated geometries, and hence this process is used throughout our current work using modified versions of the \textit{VAorthog} and \textit{VAeval} codes presented in \cite{costa2023aaa}, see their figure 11. Third, to avoid numerical errors known as Froissart doublets (see e.g. \cite{nakatsukasa2018aaa}), poles with sufficiently small residue below $\mathcal{O}(10^{-8})$ and those sufficiently close to the boundary (within $\mathcal{O}(10^{-2})$) are manually removed.

\section{Multi-domain nAAALS method}\label{sec:NdomAAALS}
Next, we extend the AAALS method described in section \ref{sec:onetwodom} to more general multi-domain problems. Consider an unbounded exterior region $\Omega_0$ punctured by $N$ simply connected regions $\Omega_n, n\in\{1,2,\dots,N\}$. Each region $\Omega_n$ is enclosed by a Jordan curve $\gamma_n$ with outward unit normal $\hat{n}_n$ and has conformal radius $L_n$. The function $\psi$ is defined everywhere and is denoted as $\psi=\psi_n$ in region $\Omega_n$, satisfying the Laplace equation in $\Omega_0$ and a Poisson equation with constant right hand side of $\omega_n$ in each interior region $\Omega_n$. The system is non-dimensionalised as in section \ref{sec:onetwodom} using the quantities $L$ and $\omega$; we typically choose $L=L_1$ and $\omega=\omega_1$ in this work. Applying the same matching Dirichlet \eqref{Meq:dBC} and Neumann \eqref{Meq:nBC} boundary conditions at each boundary $\gamma_n$ and once again imposing the far-field condition $\psi_0\rightarrow \tilde{G}_0(z)$ gives the following dimensionless system
\begin{gather}
    \nabla^2\psi_0=0\;\;\;\text{ in $\Omega_0$,}\label{Meq:mpeGE}\\
    \nabla^2\psi_n=\sigma_n\;\;\;\text{ in $\Omega_n$,}\label{Meq:mpiGE}\\
    \psi_0=\psi_n\;\;\;\text{ on $\gamma_n$,}\label{Meq:mpdBC}\\
    \frac{\p \psi_0}{\p \hat{n}_n} = \mu_n\frac{\p\psi_n}{\p\hat{n}_n}\;\;\;\text{ on $\gamma_n$,}\label{Meq:mpnBC}\\
    \psi_0\rightarrow \tilde{G}_0(z)\;\;\;\text{as $r\rightarrow\infty$,}\label{Meq:mpfar}
\end{gather}
where $\sigma_n = \omega_n/\omega$ and $\mu_n$ are constants.

The nAAALS method is developed to solve for the system \eqref{Meq:mpeGE}-\eqref{Meq:mpfar}. Each function $\psi_n$ can be written as $\psi_n=\Real[G_n(z)]+\Real[F_n(z)]$ for all $n\in\{0,1,\dots,N\}$, where the function $G_0(z)$ accounts for the far-field condition, noting that $\tilde{G}_0(z)=\Real[G_0(z)]$ as before. The additional functions $G_n(z)$ for $n\in\{1,2,\dots,N\}$ account for the non-zero right-hand sides of \eqref{Meq:mpiGE} and are 
\begin{equation}
    G_n(z) = \frac{\sigma_n}{4}|z-C_n|^2,\label{Meq:Gn}
\end{equation}
where $C_n$ is the dimensionless centre of region $\Omega_n$. The remaining terms `$\Real[F_n(z)]$' are harmonic and decay in the far-field, where $F_n(z)$ is analytic in region $\Omega_n$ and can be approximated as
\begin{gather}
    F_0(z) = \sum_{n=1}^N\Bigg[\delta_n\log{\bigg(\frac{z-C_n}{z-C_{n'}}\bigg)}+\sum_{m=1}^{M}\alpha_{nm}(z-C_n)^{-m}+\sum_{m=1}^{M_{0n}}\frac{\beta_{nm}}{z-q_{nm}}\Bigg],\label{Meq:mppsie}\\
    F_n(z) = \sum_{m=0}^{M}a_{nm}(z-C_n)^m+\sum_{m=1}^{M_{n}}\frac{b_{nm}}{z-p_{nm}}\label{Meq:mppsii},
\end{gather}
where $n'=n \text{ (mod N)}+1$. The interior functions $F_n(z)$ for $n\in\{1,2,\dots,N\}$ follow the same structure as \eqref{eq:onedomrat} where $M=20$ is the power series truncation and $p_{nm}$ are the $M_n$ poles exterior to region $\Omega_n$. Following \cite{costa2023aaa}, the exterior function $F_0(z)$ requires additional log terms to account for the $N$ punctures in $\Omega_0$, with the additional (implicit) constraint that $\sum_{n=1}^N\delta_n=0$ to ensure non-singular behaviour in the far-field. Furthermore, $F_0(z)$ is composed of $N$ power series and simple pole terms to account for each region $\Omega_n$ centred at $C_n$ and the $M_{0n}$ poles $q_{nm}$ inside each region $\Omega_n$. The poles $(p_{nm}$, $q_{nm})$ are found by using the AAA algorithm to approximate the function $\tilde{S}(z)$ that satisfies $\tilde{S}(Z)=\overline{Z}$, where $Z$ is now the list of boundary data for all curves $\gamma_n$. The function $\tilde{S}(z)$ is discussed more in section \ref{sec:discussion}, however the main point to note is that the resulting poles account both for each individual boundary geometry and the overall geometry of the $N-$domain configuration.

There are now $M_T = 2[N(2M+1)+M_{0n}+M_{n}]+N$ unknowns to find: the complex coefficients $\alpha_{nm}, a_{nm}, \beta_{nm}$ and $b_{nm}$, and the real coefficients $\delta_n$. As in section \ref{sec:onetwodom}, \eqref{Meq:mppsie} and \eqref{Meq:mppsii} can be substituted into the boundary conditions \eqref{Meq:mpdBC} and \eqref{Meq:mpnBC} and rearranged to give
\begin{gather}
        \Real[F_0(z)]-\Real[F_n(z)] = H_{Dn}(z) \;\text{ on }\gamma_n, \label{eq:mpcdBC}\\
        \Real[\hat{n}_nF_0'(z)] - \mu_n\Real[\hat{n}_nF_n'(z)] = H_{Nn}(z)\;\text{ on }\gamma_n, \label{eq:mpcnBC}
\end{gather}
where the functions $H_{Dn}$ and $H_{Nn}$ are defined as
\begin{gather}
    H_{Dn}(z) = -\Real[G_0(z)]+\Real[G_n(z)], \\
    H_{Nn}(z) = -\Real[\hat{n}_nG_0'(z)] + \mu_n\Real[\hat{n}_nG_n'(z)].
\end{gather}
Equations \eqref{eq:mpcdBC} and \eqref{eq:mpcnBC} can be written as the product $Ac=H$, where $A$ is the matrix of Vandermonde with Arnoldi basis vectors (see section \ref{sec:onetwodom}), $c$ is the unknown coefficient vector\footnote{The size of $c$ is $(M_T+N)\times1$ as the entry $\Imag[\delta_n]=0$ must also be included for each $n\in\{1,2,\dots,N\}$.}, and $H$ is the solution vector $H = (H_{D1}(Z|_{\gamma_1})\;\;H_{N1}(Z|_{\gamma_1})\;...\;H_{DN}(Z|_{\gamma_N})\;\;H_{NN}(Z|_{\gamma_N}))$. Selecting $P$ points around each boundary $\gamma_n$ allows us to solve for the $M_T$ unknown coefficients via the least-squares fitting $c=H\backslash A$. 


Once again, the nAAALS method uses the same framework as the original one-domain method from \cite{costa2023aaa}. The differences are that the AAA algorithm now outputs poles that are adapted to both the individual boundary geometries and to the overall multi-domain configuration, and the least-squares fitting operates on the matrix product $Ac=H$, which grows linearly with the number of domains added. Therefore, the computational cost also increases linearly; for example, applying the nAAALS method to a three-domain problem will require twice the computation time than that of an analogous two-domain problem. While adding additional domains increases the computation time, the cost is still comparatively small, typically running from seconds to tens of seconds for problems with many domains.

\section{Multi-domain vortex patch problems}\label{sec:multidom}
\begin{figure}[t]
    \centering
    \includegraphics[width=0.99\linewidth]{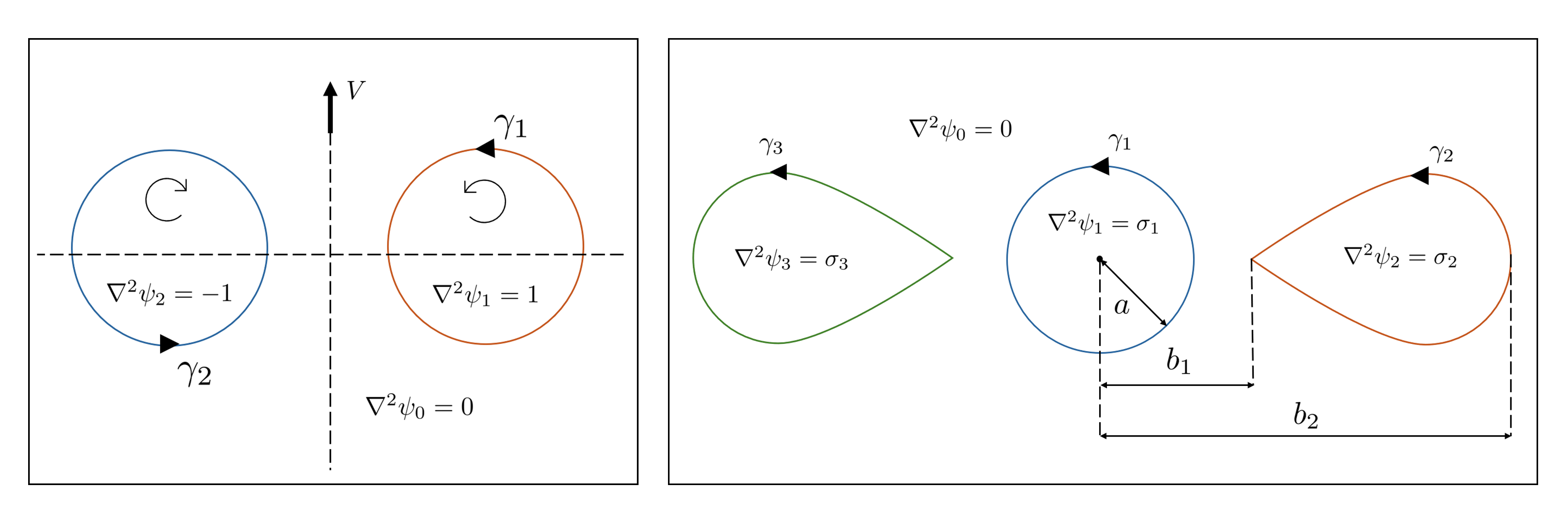}
    \caption{Mathematical diagram of two, dimensionless configurations of vortex patches. Left: a translating, counter-rotating vortex patch pair, where dotted lines denote the lines of symmetry in the problem. Right: rotating, multipolar vortex patches; this example shows a central vortex patch of radius $a$ centred at the origin surrounded by two symmetric satellite patches. The quantities $b_1$ and $b_2$ are defined as the inner and outer satellite distances, respectively.}
    \label{fig:tr_vortex_patch}
\end{figure}

To illustrate the nAAALS method developed in section \ref{sec:NdomAAALS}, we apply it to some example problems in vortex dynamics. In particular, consider a configuration of $N>0$ distinct vortex patches, finite-area regions $\Omega_n$ of fluid in rotational flow with constant vorticity $\omega_n$, surrounded by the same incompressible, inviscid fluid in irrotational flow in the exterior region $\Omega_0$. While the vorticity jumps from zero to $\omega_n$ across each patch boundary $\gamma_n$, the stream function and velocity are continuous everywhere. Non-dimensionalising the system by given length and vorticity scales $L$ and $\omega$, respectively, the stream function $\psi$ thus satisfies the system of equations \eqref{Meq:mpeGE}-\eqref{Meq:mpfar} when setting $\mu_n=1$ for all $n$, noting that the far-field function $\tilde{G}_0(z)$ depends upon the specific vortex patch problem considered. Hence the stream function can be computed numerically using the nAAALS method and the velocity vector (in complex notation) is found given the relation $u+i v = i\nabla\psi=-\psi_y+i\psi_x$. In this section, we seek vortex equilibria, configurations of these $N$ vortex patches that steadily translate or rotate over time. Two such vortex configurations are considered: a translating, counter-rotating vortex patch pair (figure \ref{fig:tr_vortex_patch}, left) and a system of $N$ co-rotating vortex patches (figure \ref{fig:tr_vortex_patch}, right).

\subsection{Translating vortex patch pair}\label{subsec:tpatches}
Consider the case of a translating, symmetric, counter-rotating vortex patch pair (figure \ref{fig:tr_vortex_patch}, left). This problem was originally presented in \cite{pierrehumbert1980family}, but here we follow closely the related work of \cite{wu1984steady}. Each vortex patch has the same representative (conformal) radius $L_1=L_2=L$ and area $A$, and both patches are centred along the real axis with dimensional centres $z=C_1^\ast$ and $z=C_2^\ast=-C_1^\ast$. Both patches have equal and opposite vorticity of representative magnitude $\omega$, hence in \eqref{Meq:mpiGE} we have that $\sigma_1=1$ and $\sigma_2=-1$. The patches translate steadily in the negative imaginary direction with constant speed $V$; upon changing the frame of reference, this can be represented by a uniform stream in the positive imaginary direction. Hence the far-field condition \eqref{Meq:mpfar} uses $\tilde{G}_0(z) = \lambda x$ where $\lambda = V/(\omega L)$.

\begin{figure}[t]
    \centering
    \includegraphics[width=0.8\linewidth]{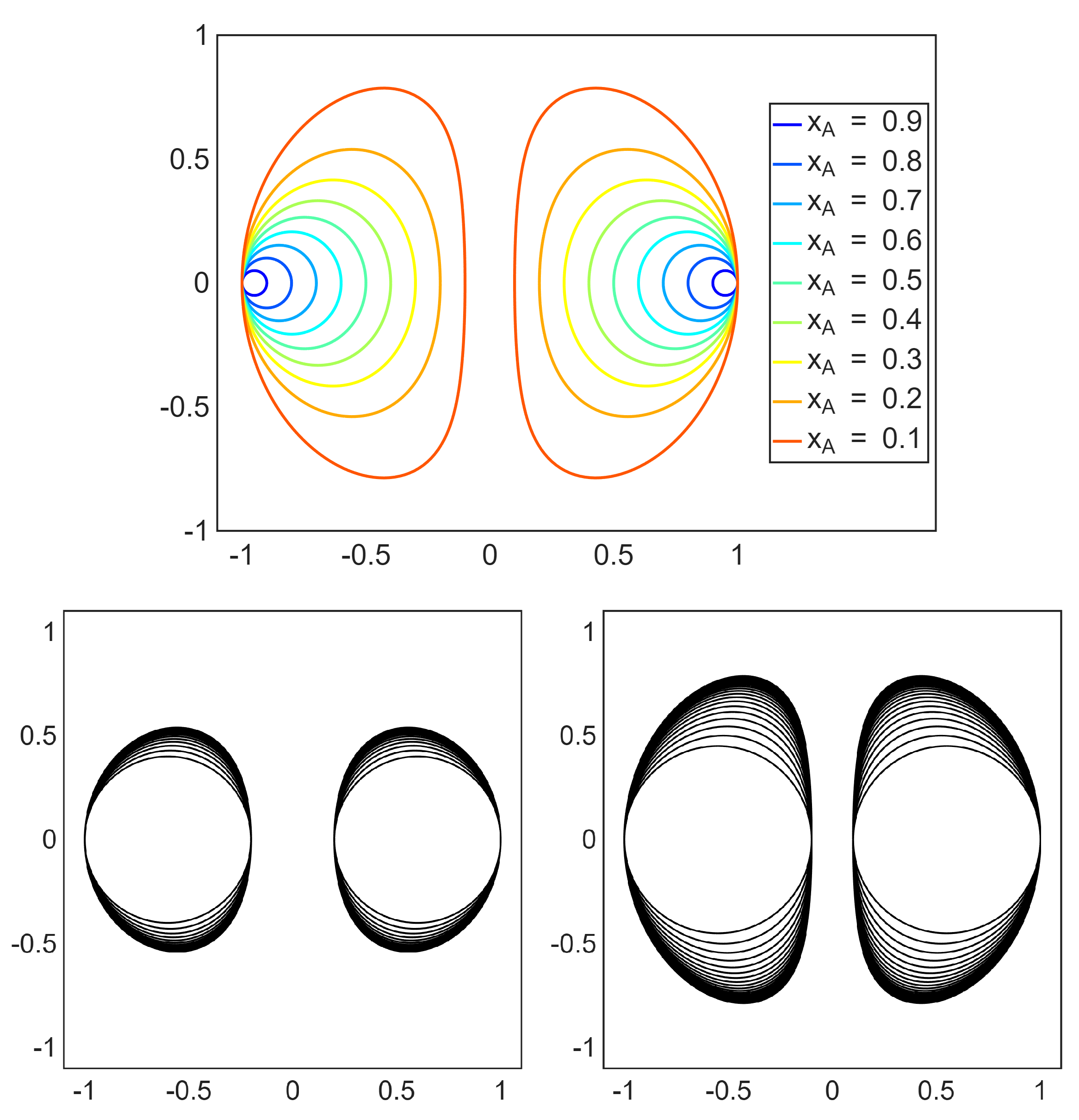}
    \caption{Top: superposition of nine translating vortex patch pair equilibria. Each right hand vortex patch has minimum and maximum $x$ distances $x_A$ and $x_B=1$, where the values of $x_A\in\{0.1,0.2,\dots,0.9\}$ for each equilibria are given in the figure legend. Bottom: iterations of the nAAALS + WOZ algorithm in computing the translating equilibria for $x_A=0.2$ (left) and $x_A=0.1$ (right).}
    \label{fig:WOZfigs}
\end{figure}

We couple the nAAALS algorithm to the numerical, iterative procedure presented by the authors Wu, Overman and Zabusky (WOZ) in \cite{wu1984steady} (details of this are given in appendix \ref{app:tWOZ}) to reproduce known vortex equilibria. For all equilibrium solutions considered, we use an initially circular vortex patch shape with radius $R=0.5(x_A+x_B)$ and centre $C_1^\ast=x_A+R$, where $z=x_A$ and $z=x_B$ are the start and endpoints of one half of the right vortex patch, respectively. We compute nine such vortex equilibria for the cases $x_A\in\{0.1,0.2,\dots,0.9\}$, with $x_B=1$ fixed in each case and $P=120$ points selected on each boundary segment. These parameters are the same as those used in \cite{wu1984steady} when computing the translating equilibria in cases 1-9 of their table I. Figure \ref{fig:WOZfigs} (top) presents the final vortex patch shapes of these equilibria, which are in qualitative agreement with the corresponding shapes given in figure 4a of \cite{wu1984steady}. Figure \ref{fig:WOZfigs} (bottom) shows a superposition of the iterations for two of these examples, specifically $x_A=0.2$ (left) and $x_A = 0.1$ (right). Convergence is achieved in under 90 iterations and a total of 20 seconds of runtime for all cases computed using the nAAALS + WOZ algorithm. While the results in \cite{wu1984steady} were computed in less than 70 iterations, each iterative step required 8 seconds of runtime. This would result in maximum total runtime of 560 seconds, demonstrating a significant speed-up when using the nAAALS + WOZ method. 

In addition to the shape of the vortex patches, the following quantities are also calculated: the area $A$, perimeter $P$ and $x$-coordinate $\bar{x}$ of the centre of the right patch; the translational speed $V$ and normalised translational speed $V/V_0$, where $V_0 = A/(4\pi\bar{x})$ is the translational speed of two line vortices with separation $2\bar{x}$ and circulation $\pm A$ \citep{wu1984steady}; the effective $\bar{R}=\sqrt{(A/\pi)}$ and normalised  $\bar{R}/\bar{x}$ radius of the right patch; and the maximum vertical extent $H=\text{max}|y|$ and aspect ratio $a=2H/(1-x_A)$ of the right patch. These values are then compared to those presented in table I of \cite{wu1984steady} by calculating the following mixed absolute-relative error
\begin{equation}\label{Meq:mixederror}
    \text{mixed error} = 
    \begin{cases}
        \frac{|x_a-x_e|}{x_{e}},\;\;|x_e|>\tau,\\
        |x_a-x_e|,\;\;|x_e|<\tau,
    \end{cases}
\end{equation}
where the values from \cite{wu1984steady} are taken as the exact (true) values $x_e$ and those computed by nAAALS + WOZ as the approximate values $x_a$. The error \eqref{Meq:mixederror} then returns the relative error if the magnitude of the exact value is above some tolerance $\tau$, and the absolute error otherwise to avoid singular behaviour when $x_e$ is close to zero; in this work, $\tau=10^{-4}$ is chosen. The nAAALS + WOZ results agree with those of \cite{wu1984steady} to at least an $\mathcal{O}(10^{-3})$ mixed error, noting that the results in table I of \cite{wu1984steady} are given to three significant figures.


\subsection{Rotating multipolar vortex patches}\label{subsec:rpatches}
Next, we consider the case of a system of multipolar, co-rotating vortex patches (figure \ref{fig:tr_vortex_patch}, right), once again with the aim of computing vortex equilibria. In particular, we reproduce numerical equilibria of two problems, first of the multipolar configurations found in \cite{dritschel1985stability} involving $N$ symmetric `satellite' patches each with two-fold symmetry and of equal vorticity, and second the `$N+1$' vortex equilibria of \cite{xue2017new} where the $N$ satellite patches surround a central patch of $N-$fold symmetry with equal, same-signed vorticity. As in section \ref{subsec:tpatches}, we can reproduce known numerical equilibria to both problems using a modified version of the WOZ algorithm coupled with the nAAALS method from section \ref{sec:NdomAAALS} - see appendix \ref{app:mrWOZ} for the details. 

The representative length $L$ and vorticity $\omega$ scales used are the corresponding quantities of each satellite patch for satelitte-only equilibria, and of the central $N-$fold symmetric patch for $N+1$ vortex equilibria. Neglecting arbitrary additive constants, the far-field function $\tilde{G}_0(z)$ in \eqref{Meq:mpfar} is given by
\begin{equation}\label{Meq:mpG0}
    \tilde{G}_0(z) = \sum_{n=1}^{N^{\ast}}\sigma_n\lambda_n\log{|z-C_n|},
\end{equation}
where $\sigma_n = \omega_n/\omega=1$ for all $n$, $\lambda_n=A_n/(2\pi L^{2})$ where $A_n$ is the (dimensional) area enclosed by $\gamma_n$, and $N^{\ast}$ is equal to $N$ for the satellite-only problem and $N+1$ for the satellites plus central patch problem. Here, $P=250$ points are selected around each $1/N$ central patch segment and each half of the satellite patch segment.

\begin{figure}[t]
    \centering
    \includegraphics[width=0.75\linewidth]{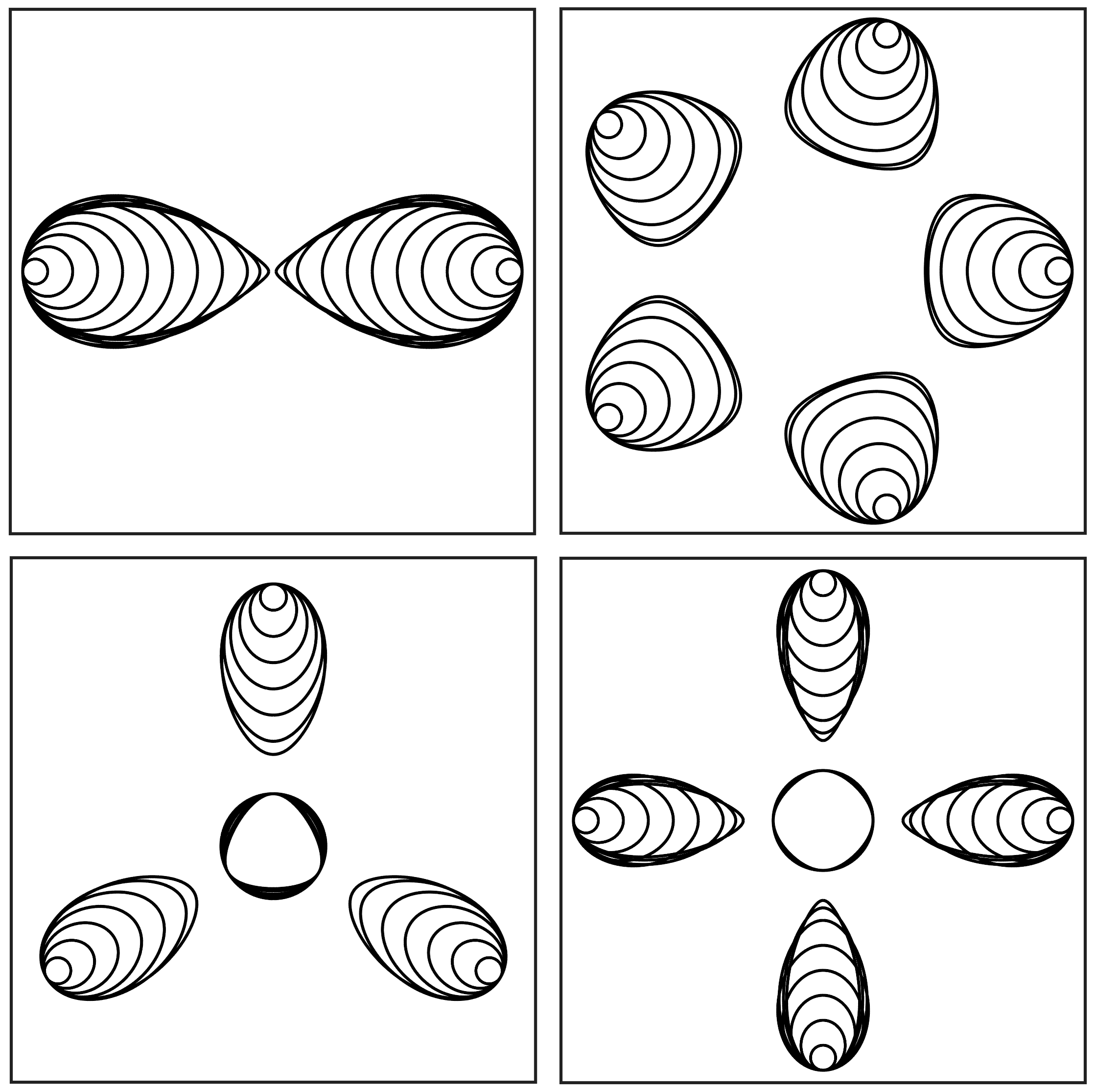}
    \caption{Superpositions of rotating, multipolar vortex patch equilibria in two-, three-, four- and five-fold symmetries (from top-left, travelling counter-clockwise to top-right). Each equilibrium uses dimensional outer satellite distance $b_2=1$ and central patch radius $a=0.2$ (where applicable). Top row: satellite-only equilibria. The two-fold example (left) shows equilibria for inner satellite distances $b_1=[0.015,0.05,0.1,0.2,0.3,0.4,0.5,0.6,0.7,0.8,0.9]$, and the five-fold example (right) for distances $b_1 = [0.44, 0.5, 0.6,0.7,0.8,0.9]$. Bottom row: central + satellite equilibria. The three-fold example (left) uses distances $b_1=[0.35,0.4,0.5,0.6,0.7,0.8,0.9]$, and the four-fold example (right) uses distances $b_1=[0.32,0.35,0.4,0.5,0.6,0.7,0.8,0.9]$.}
    \label{fig:XDfigs}
\end{figure}

Figure \ref{fig:XDfigs} shows examples of two-, three-, four- and five-fold symmetric equilibria computed using the nAAALS + WOZ algorithm. Each figure gives a superposition of multiple $n-$fold symmetric equilibria for different values of the inner satellite distance $b_1$ (see the figure caption for these ranges), while the dimensional central patch radius and outer satellite distance are fixed as $a=0.2$ and $b_2=1$, respectively. Geometric interpretations of these distances are presented in figure \ref{fig:tr_vortex_patch} (right). The two- and five-fold examples (top row) recreate the examples from figures 2a and 2d of \cite{dritschel1985stability}, respectively, while the three- and four-fold examples (bottom row) recreate figures 7a and 7b from \cite{xue2017new}, with all figures showing good qualitative agreement with the original figures they are reproducing. 

Additionally, we calculated the normalised areas, angular velocities and centroid positions of the satellite patches in the two- and five-fold equilibria of figure \ref{fig:XDfigs} and computed the mixed error \eqref{Meq:mixederror} using the results from table 1 of \cite{dritschel1985stability} as the exact values. There was generally agreement of $\mc{O}(10^{-5})$ for all computed values, except for the limiting configuration in the two-fold symmetric case where there was only agreement to $\mc{O}(10^{-4})$. Note that the results listed in table 1 of \cite{dritschel1985stability} are only given to 5 decimal places, indicating why we may be unable to achieve an error smaller than $\mc{O}(10^{-5})$.

\section{Nested domain problems}\label{sec:nested}
In this section, we consider another multi-domain scenario known as a nested domain problem. Unlike the multi-domain problems considered in section \ref{sec:multidom} where the exterior domain $\Omega_0$ is punctured by $N$ simply connected domains, here $\Omega_0$ is punctured by one domain which itself is punctured by one or more domains. As such, the domains can be thought of as `stacked' or `nested' within each other. 

\begin{figure}[t]
    \centering
    \includegraphics[width=0.99\linewidth]{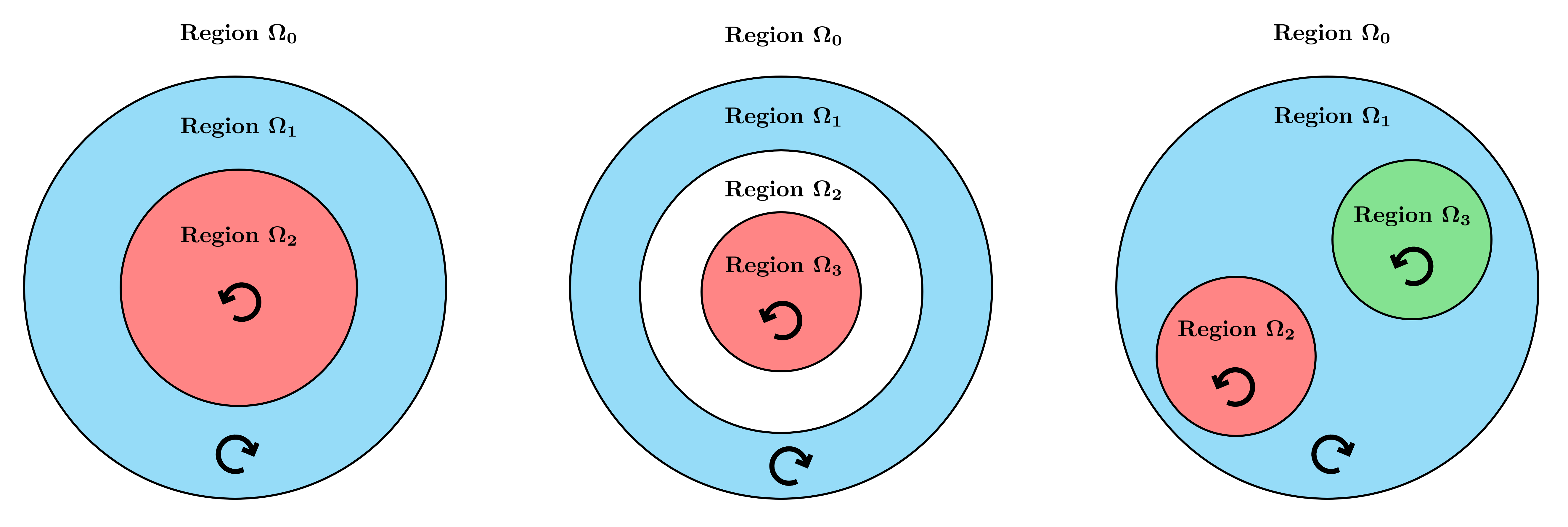}
    \caption{Examples of nested vortex patches; in all cases, region $\Omega_0$ represents the exterior region of irrotational flow. Left: the two-contour Rankine vortex, consisting of a central core (region $\Omega_2$) of vorticity $\gamma<0$ and an outer annulus (region $\Omega_1$) of positive unit vorticity. Centre: the three-contour Rankine vortex, similar to the two-contour vortex, however includes an additional interior irrotational annulus (now region $\Omega_2$). Right: a cookie vortex, consisting of a primary vortex patch (region $\Omega_1$) of unit vorticity punctured by two smaller vortex patches (regions $\Omega_2$ and $\Omega_3$) of vorticity $\gamma<0$.}
    \label{fig:N-contour_Rankine_diagram}
\end{figure}

To illustrate the application of the nAAALS method to these nested domain problems, two classes of problem are considered. First is the class of stacked problems that consist of $N$ domains of decreasing conformal radii. Each domain (apart from the central, smallest region) is punctured by exactly one smaller domain, which itself is punctured by a smaller domain, and so on. Two example configurations are given in figure \ref{fig:N-contour_Rankine_diagram} (left and centre) in the context of vortex patches, where all patches are centred at the origin. In particular, these are the two- (left) and three-contour (centre) Rankine vortices, which are known nested vortex patch equilibria, see section \ref{subsec:nestedresults}. Second is the class of `cookie' problems, where a primary domain is punctured by many smaller, non-punctured domains, an example configuration is given in figure \ref{fig:N-contour_Rankine_diagram} (right).

\subsection{Nested domain nAAALS algorithm}
The non-dimensional mathematical model and AAALS method used for nested problems are similar to the nAAALS method in section \ref{sec:NdomAAALS}, however the location of the boundary conditions and the form of the rational approximations differ. The governing equations \eqref{Meq:mpeGE} and \eqref{Meq:mpiGE} hold, as does the far-field condition \eqref{Meq:mpfar}. The functions $\psi_n$ for $n\in\{0,1,\dots,N\}$ are written as $\psi_n = \Real[G_n(z)]+\Real[F_n(z)]$, where $\tilde{G}_0(z)=\Real[G_0(z)]$ has the form of \eqref{Meq:mpG0} for all problems considered in this section and the remaining $G_n(z)$ by \eqref{Meq:Gn}. The functions $F_n(z)$ are largely the same as in \eqref{Meq:mppsie} and \eqref{Meq:mppsii}, however with notable differences. There is now no logarithm term in the expression for $F_0(z)$ as the exterior region contains only a single puncture at the origin. For the remaining regions, additional logarithm  `$\log(z-C_k)$' and reciprocal `$1/(z-C_k)^m$' terms must be included in the expressions $F_n(z)$ to account for each $k$ puncture (if any) within the region \citep{costa2023aaa}. Therefore,
\begin{equation}
    F_0(z) = \sum_{m=1}^{M}a_{0m}(z-C_1)^{-m}+\sum_{m=1}^{M_{0}}\frac{b_{0m}}{z-p_{0m}},
\end{equation}
\begin{equation}
        F_n(z) = \sum_{m=0}^{M}a_{nm}(z-C_n)^m+\sum_{m=1}^{M_{n}}\frac{b_{nm}}{z-p_{nm}} + \sum_{k=1}^K\bigg(\sum_{m=1}^M c_{nmk}(z-C_k)^{-m} + d_{nk}\log(z-C_k) \bigg),
\end{equation}
where $K\geq0$ is the total number of punctures in region $n\in\{1,2,\dots,N\}$. Once again, $p_{nm}$ are the $M_n$ poles exterior to the region $\Omega_n$ and are found by approximating the function $\tilde{S}(z)$ where $\tilde{S}(Z)=\overline{Z}$ on the list $Z$ of all boundary data. The unknown complex coefficients $a_{nm}$, $b_{nm}$, $c_{nmk}$ and the real coefficients $d_{nk}$ are then to be found.

The boundary conditions are slightly different to those used in section \ref{sec:NdomAAALS}, and are now
\begin{gather}
    \psi_{n^\ast}=\psi_{n}\;\;\;\text{ on $\gamma_n$,}\label{Meq:npdBC}\\
    \frac{\p \psi_{n^\ast}}{\p \hat{n}_n} = \frac{\p\psi_n}{\p\hat{n}_n}\;\;\;\text{ on $\gamma_n$.}\label{Meq:npnBC}
\end{gather}
Here, the index $n^\ast<n$ is the index of the region neighbouring region $\Omega_n$. In the stacked problem, $n^\ast=n-1$, while in the cookie problem $n^\ast = 0$ when $n=1$ and $n^\ast = 1$ for all other $n>1$. These boundary conditions can then be written in terms of the functions $F_n(z)$ and $G_n(z)$ as
\begin{gather}
    \Real[F_{n^\ast}(z)] - \Real[F_{n}(z)] = -\Real[G_{n^\ast}(z)]+\Real[G_{n}(z)]\;\text{ on }\gamma_n,\label{eq:npcdBC}\\
    \Real[\hat{n}_nF_{n^\ast}'(z)] - \Real[\hat{n}_nF_{n}'(z)] = -\Real[\hat{n}_nG_{n^\ast}'(z)] + \Real[\hat{n}_nG_{n}'(z)]\;\text{ on }\gamma_n.\label{eq:npcnBC}
\end{gather}
Equations \eqref{eq:npcdBC} and \eqref{eq:npcnBC} can be written as the product $Ac=H$ and the unknown coefficients solved in the usual least-squares way.

\subsection{Stacked and cookie vortex patches}\label{subsec:nestedresults}
The nested domain variant of the nAAALS method is now used to generate various nested vortex patches in both stacked and cookie configurations. First, to test the accuracy of the algorithm, its solutions are compared to two existing analytic stacked vortex patch equilibria. These are the two- and three-contour Rankine vortices displayed in figure \ref{fig:N-contour_Rankine_diagram} (left and centre, respectively), both of which are presented in \cite{morel1994multipolar}. The two-contour Rankine vortex consists of a central, circular region of radius $B$ and vorticity $\gamma<0$ surrounded by an annular region between $|z|=B$ and $|z|=A_2$ of positive unit vorticity, and an irrotational region for $|z|>A_2$. The three-contour vortex is similar, but with an additional annular region of irrotational flow between $|z|=B$ and $|z|=A_1<A_2$; the region of positive unit vorticity is then between $|z|=A_1$ and $|z|=A_2$. Analytical solutions for the velocity field of the two- and three-contour Rankine vortices are given in appendix \ref{app:analy}.

\begin{figure}[t]
    \centering
    \includegraphics[width=0.9\linewidth]{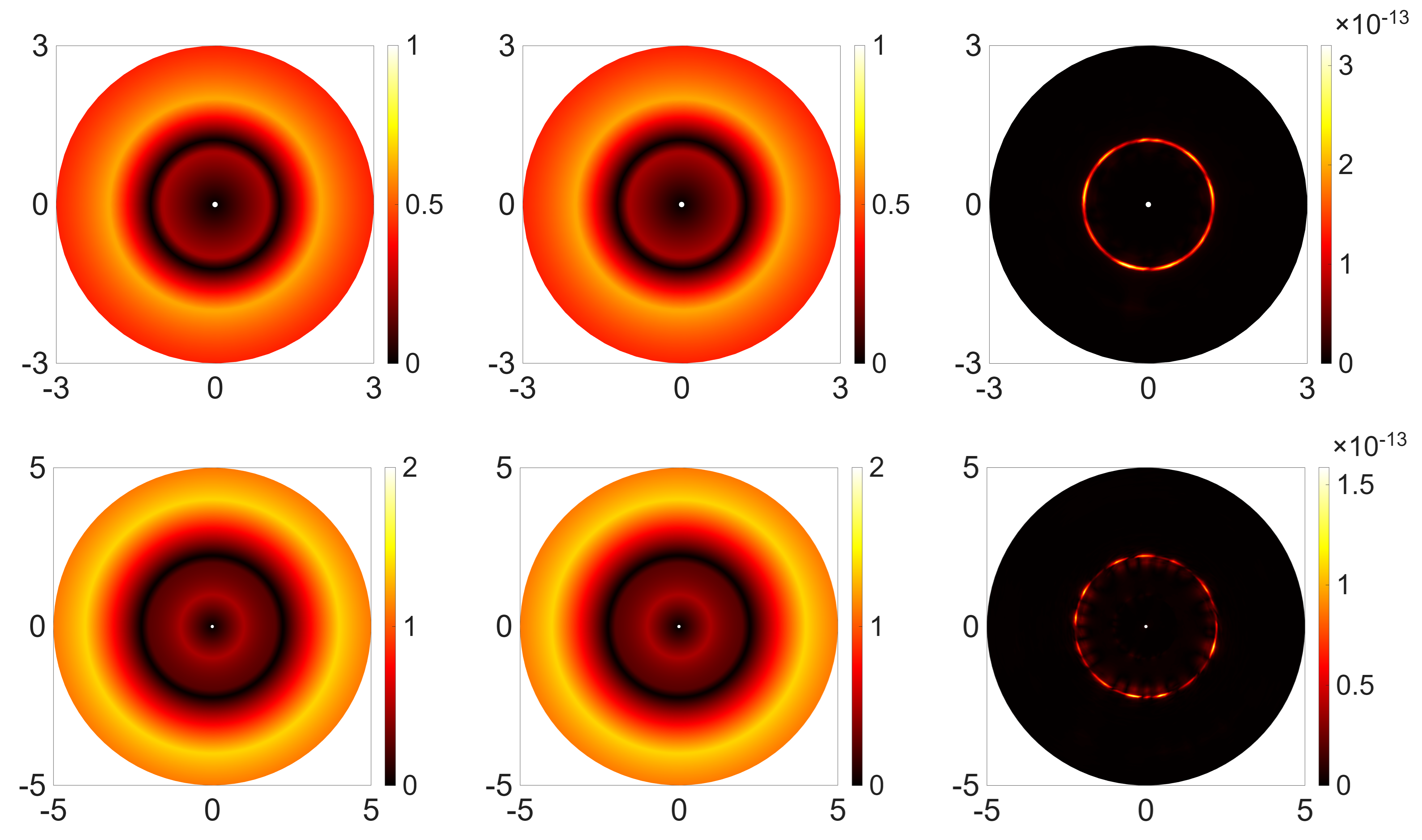}
    \caption{Contour plots of the nAAALS (left) and exact (centre) velocity magnitudes of a two-contour Rankine vortex with $\gamma=-0.5$, $B=1$ and $A_2=2$ (top) and a three-contour Rankine vortex with $\gamma=-1$, $B=1$, $A_1=2$ and $A_2=4$ (bottom). Right: contour plots of the mixed error between the AAALS and exact solutions.}
    \label{fig:N-contour_Rankine}
\end{figure}

Figure \ref{fig:N-contour_Rankine} gives two examples of these Rankine vortices: the top row presents the nAAALS (left) and exact (centre) velocity magnitudes for a two-contour vortex with $\gamma=-0.5$, $B=1$ and $A_2=2$; similarly, the bottom row gives solutions for a three-contour vortex with $\gamma=-1$, $B=1$, $A_1=2$ and $A_2=4$. The right column then gives a contour plot of the mixed error \eqref{Meq:mixederror} of the nAAALS result compared to the exact solution, which is found to be of $\mathcal{O}(10^{-13})$. Note that all nested solutions found using the nAAALS method in this section were computed once again on the order of seconds of runtime.

\begin{figure}[t]
    \centering
    \includegraphics[width=0.99\linewidth]{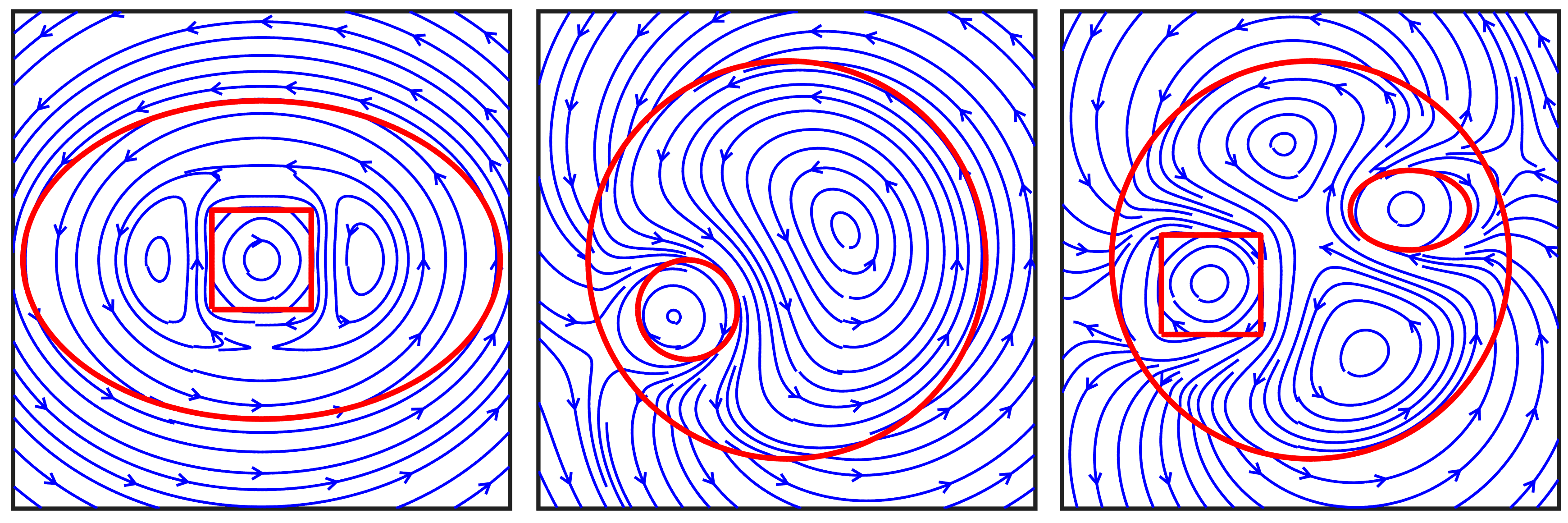}
    \caption{Examples of dimensional, non-equilibrium nested vortex patches; in each case, the vortex patch boundaries are displayed in red and the streamlines in blue, with arrows indicating the direction of flow. Left: two-contour vortex patch, with both patches centred at the origin. Outer ellipse has major and minor axes $a=4.8$, $b=3.2$ and unit vorticity; inner square has side length $l=2$ and vorticity $\gamma=-2$. Centre: two-contour Rankine vortex with off-centre inner patch. Outer circle has radius $r=4$ and unit vorticity; inner circle has radius $r=1$, is centred at $z=-2-i$, and has vorticity $\gamma=-5$. Right: cookie vortex patch. Outer circle has radius $r=4$, is centred at the origin, and has unit vorticity. Left puncturing square patch has side length $l=2$, is centred at $z=-2-0.5i$, and has vorticity $\gamma=-4$. Right puncturing elliptic patch has major and minor axes $a=1.2$ and $b=0.8$, is centred at $z=2+i$, and has vorticity $\gamma=-4$.}
    \label{fig:nested_examples}
\end{figure}

Figure \ref{fig:nested_examples} presents dimensional examples of three non-equilibrium nested vortex patch configurations whose stream functions and velocities were computed with the nAAALS method. In each figure, vortex patch boundaries are displayed in red and streamlines in blue, with the arrows indicating the direction of flow. Since these patches are not in equilibrium, this is the initial shape of the configuration before the patch boundaries would be advected by the flow. The left image of figure \ref{fig:nested_examples} shows a two-contour stacked vortex patch with both patches centred at the origin. The outer patch is an ellipse with major and minor axes $a=4.8$ and $b=3.2$ and unit vorticity, while the inner patch is a square with side length $l=2$ and vorticity $\gamma=-2$. The centre figure shows a two-contour Rankine vortex with an off-centre inner circle. The outer patch has radius $r=4$ with unit vorticity, while the inner patch has radius $r=1$, is centred at the point $z=-2-i$ and has vorticity $\gamma=-5$. The right figure shows a cookie vortex patch. The primary patch is a circle of radius $r=4$ centred at the origin with unit vorticity. The left puncturing patch is a square with side length $l=2$, centred at $z=-2-0.5i$, and of vorticity $\gamma=-4$, while the right puncturing patch is an ellipse with major and minor axes $a=1.2$, $b=0.8$, respectively, centred at $z=2+i$, and of vorticity $\gamma=-4$. 

\begin{figure}[t]
    \centering
    \includegraphics[width=0.99\linewidth]{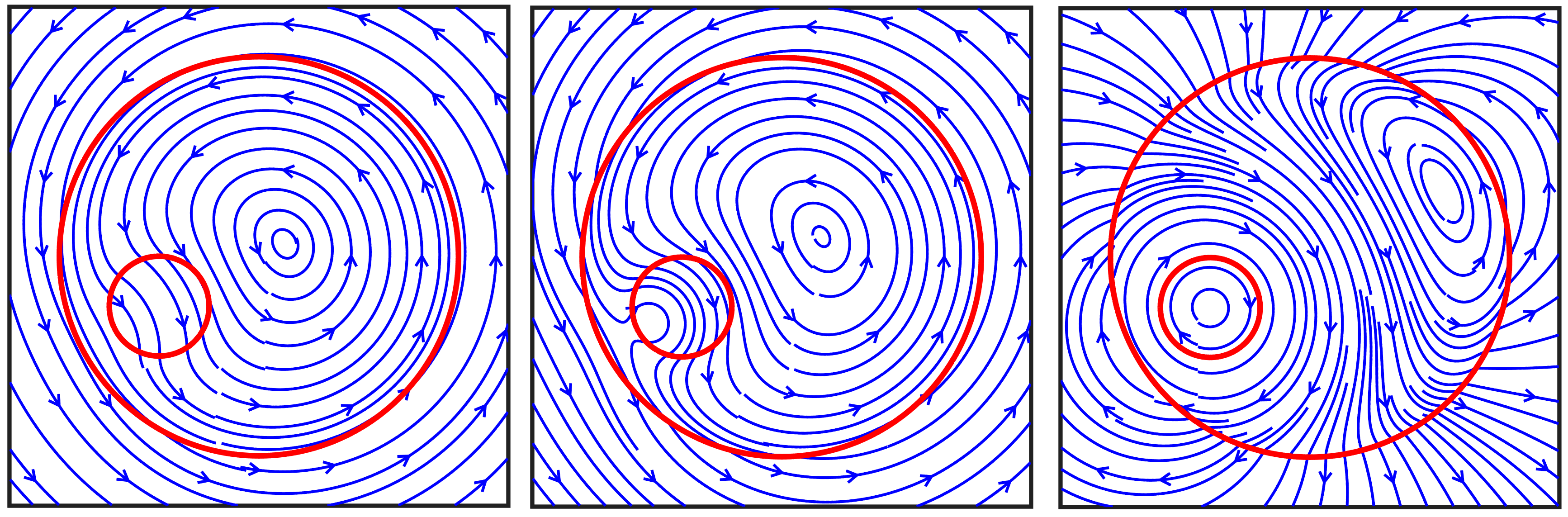}
    \caption{Further dimensional examples of the nested vortex patch configurations in figure \ref{fig:nested_examples} (centre). Outer circle has radius $r=4$ and unit vorticity; inner circle has radius $r=1$, is centred at $z=-2-i$, and has vorticity $\gamma=-1$ (left), $\gamma=-2.5$ (centre) and $\gamma=-20$ (right).}
    \label{fig:two-contour_non-centre_v1}
\end{figure}

Stagnation points develop in each example as a result of the smaller, interior vortex patches having vorticity opposite and greater in magnitude to the larger patch. Figure \ref{fig:two-contour_non-centre_v1} shows further dimensional examples of the nested vortex patch in figure \ref{fig:nested_examples} (centre) where the vorticity of the smaller patch is varied between $\gamma=-1$ (left), $\gamma=-2.5$ (centre) and $\gamma=-20$ (right). While the smaller values of vorticity $\gamma=-1,-2.5$ are not strong enough to reverse the flow, the larger values of vorticity $\gamma=-5,-20$ result in an additional stagnation point within the smaller vortex patch. Additionally, as the magnitude of the vorticity of the smaller patch increases, the stagnation point inside this patch tends towards the patch centre, while the stagnation point in the larger vortex patch is pushed further towards the patch boundary. To the authors' knowledge, the non-equilibrium nested vortex patches in this section have not been explicitly considered previously in the literature and hence we cannot compare our solutions to any existing work. In fact, the literature on nested vortex patches generally remains sparse; this is commented on in more detail in section \ref{sec:discussion}.

\section{Time-stepping problems}\label{sec:time-stepping}
The AAALS method is computationally fast, running in fractions-of-seconds to seconds on a standard laptop. While adding additional domains linearly increases the runtime of the algorithm (see section \ref{sec:NdomAAALS}), the nAAALS method still runs quickly, on the order of seconds to tens-of-seconds on a laptop. This makes the AAALS and nAAALS methods well suited to couple to time-stepping algorithms, such as the family of Runge-Kutta (RK) time-stepping methods (see e.g. \cite{dormand1980family}), to perform quasi-steady time-stepping \citep{harris2026computation,nakatsukasa2026applications} by using the nAAALS method at each time-step to iteratively solve the problem until all desired time-steps are computed. In this work, the nAAALS method is coupled to the standard RK4 time-stepping method, with the MATLAB functions `smoothdata' and `interparc' \citep{derrico2024interparc} used at each iteration to smooth numerical noise and remesh the evolved boundary curve, respectively.

In this section, the nAAALS + TS algorithm is demonstrated by applying it once again to problems in vortex dynamics, in particular to the evolution of non-equilibrium nested vortex patch configurations as introduced in section \ref{sec:nested}. First, we consider the case of a perturbed, two-contour Rankine vortex following the work of \cite{dritschel1986nonlinear}, where both patches are centred at the origin. We wish to recreate their figure 8, which considers the case where the inner patch has radius $a=0.7$ and zero vorticity, and the outer patch has unit radius and vorticity. Each boundary $\gamma_n$ is then perturbed by $0.1\cos{2(\theta-\beta_n)}$, where $\theta$ is the argument of $z$, $\beta_1=0$, and $\beta_2=\pi/2$ (see their table 2). 

\begin{figure}[t]
    \centering
    \includegraphics[width=0.85\linewidth]{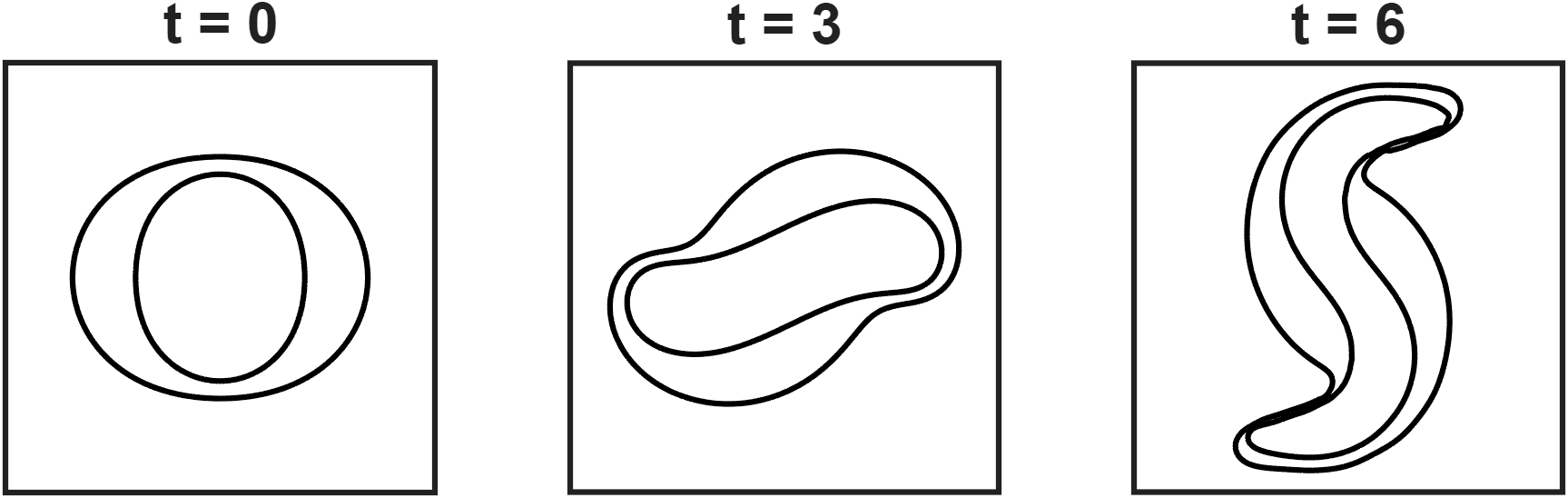}
    \caption{Evolution of a perturbed, centred, two-contour Rankine vortex. The inner vortex patch has unperturbed radius $a=0.7$ and zero vorticity; the outer patch has unit radius and vorticity. Each boundary $\gamma_n$ is perturbed by $0.1\cos{2(\theta-\beta_n)}$, where $\beta_1=0$ and $\beta_2=\pi/2$. In the results presented, time is scaled by the quantity $T=(1+a^2)/(1-a^2)$.}
    \label{fig:Dritschel_tstep}
\end{figure}

Figure \ref{fig:Dritschel_tstep} presents the evolution of the perturbed, centred, two-contour Rankine vortex, where $P=500$ points are selected on each boundary. Following \cite{dritschel1986nonlinear}, the time is scaled to the quantity $T=(1+a^2)/(1-a^2)$ and so the nested vortex patch is evolved up to a final scaled time of $t^\ast=6$ (or unscaled time of $t=17.53$). A time-step of $dt=0.05$ is used and the computation takes just under $15$ minutes of runtime. The results compare qualitatively well to those from figure 8 of \cite{dritschel1986nonlinear}, however note the presence of numerical instabilities in the upper part of the nested vortex patch in our figure \ref{fig:Dritschel_tstep} (right). This is a limitation of the current nAAALS method: at the onset of filamentation (which is observed in figure \ref{fig:Dritschel_tstep} (right)), the boundary contours lie very closely to each other. Poles are required in this region between the two contours to represent the singular behaviour there, however these poles are too close to the boundaries that results in numerical instabilities on both interfaces. Consequently, the nAAALS + TS method can only iterate to a final time of $t^\ast=6$, whereas the original results presented in \cite{dritschel1986nonlinear} could compute further, up to a final time of $t^\ast=9$. At these later times, the vortex configuration breaks up and subsequently merges again; such a splitting-merging mechanism has not been incorporated into our current nAAALS + TS method, however such a feature is feasible in a similar way to e.g. \cite{harris2025modelling}.

\begin{figure}[t]
    \centering
    \includegraphics[width=0.85\linewidth]{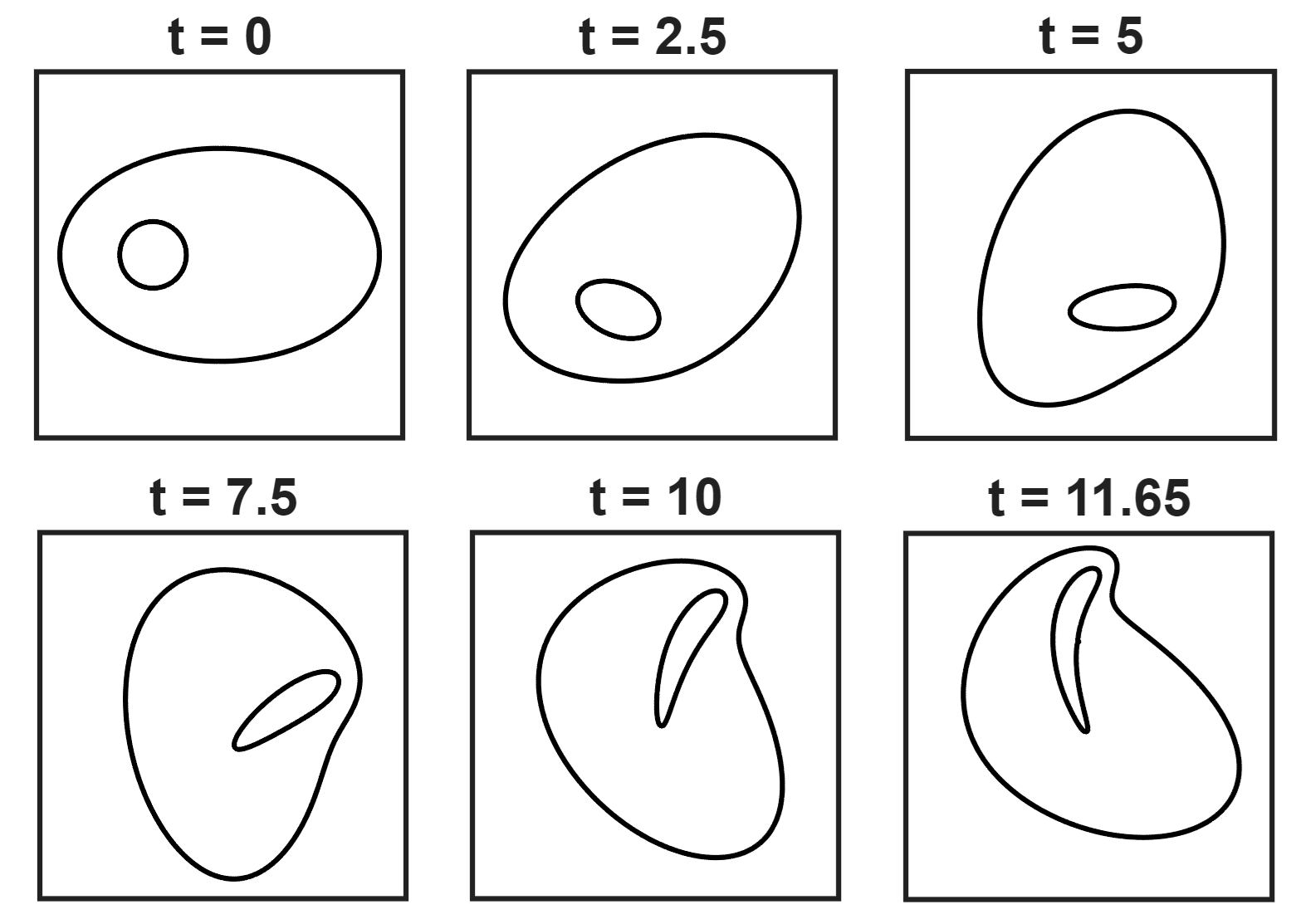}
    \caption{Evolution of a perturbed, non-centred, two-contour Rankine vortex. The inner vortex patch has unperturbed radius $a=0.5$, vorticity $\gamma=-0.5$, and is centred at $z=-1$; the outer patch has radius $a=2$, unit vorticity, and is centred at the origin. Only the outer patch is perturbed by the quantity $0.1\cos{2\theta}$.}
    \label{fig:non-centre_tstep}
\end{figure}

While \cite{dritschel1986nonlinear} does not consider cases where the inner vortex patch of the two-contour Rankine vortex is off-centre or has non-zero vorticity, we give such an example in this work that is shown in figure \ref{fig:non-centre_tstep}. The inner contour is now an unperturbed circle of radius $a=0.5$ and vorticity $\gamma=-0.5$ centred at $z=-1$; the outer contour is a circle of radius $a=2$ and unit vorticity that is centred at the origin and once again perturbed by the quantity $0.1\cos{2\theta}$. Both contours consist of $P=500$ points and a time-step of $dt = 0.05$ is used up to a maximum (unscaled) time of $t=11.76$. The computation took just under $5$ minutes of runtime and produced stable evolution of the nested vortex patch. However, by the final time $t=11.76$, the inner contour has developed a small instability that does not allow the time-stepping to progress further. Once again, it is expected that this is a consequence of the onset of filamentation.

\begin{figure}[t]
    \centering
    \includegraphics[width=0.95\linewidth]{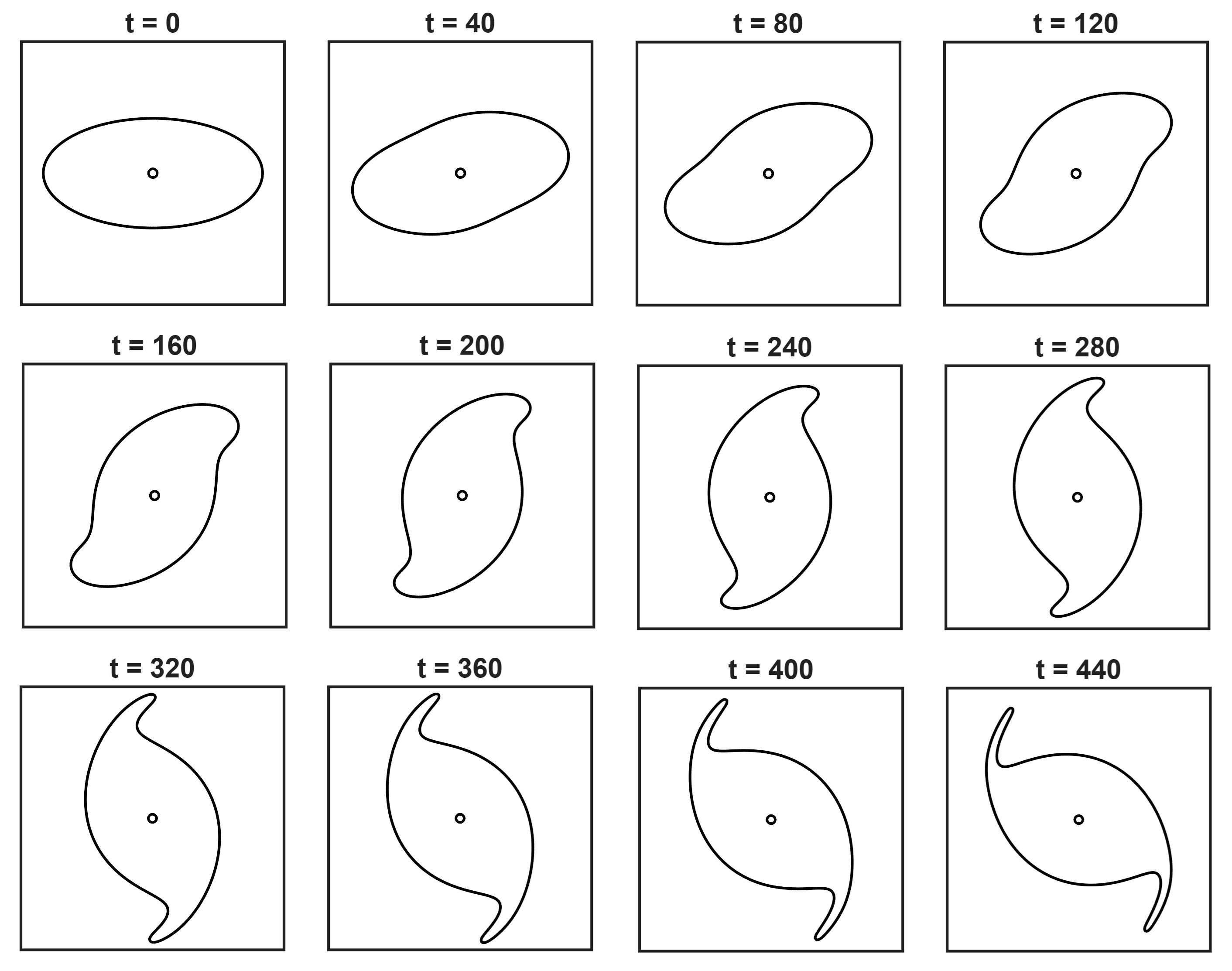}
    \caption{Dimensional evolution of a centred tornado-mesocyclone pair. The outer vortex patch (the mesocyclone) is elliptical with major semi-axis $R=2.5$km, aspect ratio $\alpha=2$, and vorticity $\omega_1=0.016\text{ s}^{-1}$; the inner patch (the tornado) is circular with radius $R=100$m, and vorticity $\omega_2=2.0\text{ s}^{-1}$.}
    \label{fig:Lin_tstep}
\end{figure}

In a vortex dynamics context, the motivation for developing such a time-stepping algorithm is to monitor and predict the development of real-world vortex patches. For example, the work of \cite{lin1992contour} considers the problem of nested vortex patches in the context of tornado-like vortices. In particular, they consider the case where a smaller tornado is embedded within a larger parent storm known as a mesocyclone; a real-world example of this is the Dallas tornado of April 2nd, 1957 \citep{hoecker1960wind}. To demonstrate the ability of the nAAALS + TS method to be used in such real-world contexts, we recreate figure 14 of \cite{lin1992contour} which presents the evolution of an (idealised) tornado-mesocyclone pair, where both tornado and mesocyclone are centred at the origin. We use the same dimensional quantities as in \cite{lin1992contour}: the outer mesocyclone is elliptical with major semi-axis $R=2.5$km, aspect ratio $\alpha=2$, and vorticity $\omega_1=0.016\text{ s}^{-1}$; the inner tornado is circular with radius $R=100$m, and vorticity $\omega_2=2.0\text{ s}^{-1}$. We select $P=1000$ points around each boundary and a time-step of $dt=0.5$s to a final time of $t=440$s. The results are given in figure \ref{fig:Lin_tstep} and were computed in just under $33$ minutes of runtime, showing excellent qualitative agreement to the original figure 14 in \cite{lin1992contour}.


\section{Discussion}\label{sec:discussion}
In this work, we have developed a multi-domain extension to the AAALS algorithm, an existing numerical method that had previously only been applied to one- \citep{costa2023aaa} and two-domain \citep{harris2025modeling} singly- and multiply-connected problems. Two variants of this nAAALS method were developed. The first considered an unbounded domain punctured by multiple, simply-connected domains, referred to here as multi-domain problems. The second was a single domain that itself contained one or more interior domains; these were referred to as nested domain problems. Furthermore, the nAAALS method was coupled to standard RK time-stepping methods, such as the RK4 method, to compute the evolution of free boundaries over time.

To assess the robustness of the proposed method, we applied it to various problems in vortex dynamics, specifically those involving vortex patches, finite-area regions of fluid in rotational flow. The nAAALS algorithm was used in conjunction with variants of the WOZ iterative procedure (see appendices \ref{app:tWOZ}, \ref{app:mrWOZ}) to reproduce known analytical and numerical vortex patch equilibria, and the coupled nAAALS + TS framework used to compute the evolution of non-equilibrium vortex patch configurations. 

As stated in section \ref{sec:introduction}, no major new results in the field of vortex dynamics are presented here, yet some novel exploration into non-equilibrium nested vortex patch configurations is introduced. Instead, the focus has been to use existing vortex dynamics results as a benchmark to demonstrate the capabilities of the nAAALS method developed in this work. While the contour dynamics techniques introduced in \cite{dritschel1989contour} remain among the most widely used numerical approaches for vortex dynamics problems, the nAAALS method is presented as an alternative framework that offers both computational efficiency and adaptability when applied to problems involving multiple vortex patches with smooth and non-smooth boundaries. Furthermore, its applicability to a broad range of problems in applied mathematics, mathematical biology, physics, and fluid mechanics makes the nAAALS method a versatile computational tool. 

There are many opportunities for further research of the nAAALS algorithm. While the method developed in this work applies only to solutions of Laplace's equation in multi-domain configurations, it is anticipated that similar extensions could also be constructed for the Helmholtz \citep{gopal2019new} and biharmonic \citep{xue2024computation} equations. Additional existing extensions to the one-domain AAALS method could also be coupled with the nAAALS method, for example using the continuum \citep{driscoll2024aaa} and/or AAA-Lawson \citep{nakatsukasa2020algorithm} versions of the AAA algorithm.

Next, it was shown in section \ref{sec:onetwodom} that for one- and two-domain problems, poles were generated using the AAA algorithm to approximate the Schwarz function $S(z)$ on the boundary $\gamma$, producing poles whose distribution was adapted to the boundary geometry. A similar method was used in the nAAALS method, where the AAA algorithm was used to approximate the function $\tilde{S}(z)$ satisfying $\tilde{S}(Z)=\overline{Z}$ on the list of all boundary data $Z$. The resulting distribution of poles was adapted both to the individual boundary geometries and to the multi-domain configuration. This new function $\tilde{S}(z)$ contains multi-domain geometrical information in a similar way to the Schwarz function for a singly-connected geometry, yet such a function has not been defined or discussed within the literature. Given the useful applications of the Schwarz function \citep{davis1974schwarz,trefethen2025numerical,harris2026computation}, future research investigating the function $\tilde{S}(z)$ is of interest.

Furthermore, there are additional problems that can be considered in the field of vortex dynamics. For example, consider the case of nested vortex patches as in section \ref{sec:nested}. Unlike the translating and rotating multipolar vortex patch problems in sections \ref{subsec:tpatches} and \ref{subsec:rpatches}, respectively, there has been little to no discussion in the literature on finding further (numerical) nested vortex patch equilibria beyond the two- and three-contour Rankine vortices. In theory, an adaptation of the WOZ algorithm like those detailed in appendices \ref{app:tWOZ} and \ref{app:mrWOZ} could be developed which iterates the shape of each vortex patch boundary until a numerical equilibrium is reached. Finding such new numerical nested vortex patch equilibria, along with a discussion on whether such solutions should be expected to exist at all, is left as an avenue for future research. 

Additionally, the nAAALS method can be used for further vortex dynamics problems not considered in this work. Such problems include computing rational approximations of the velocity field of configurations involving multiple point vortices, vortex sheets (two-dimensional line segments of vorticity) and vortex patches, as well as configurations combining the multi- (section \ref{sec:multidom}) and nested domain (section \ref{sec:nested}) scenarios considered here. Following the method developed in this work of coupling the nAAALS algorithm to the WOZ iterative scheme and to RK time-stepping methods, further numerical vortex equilibria for these new configurations could be discovered and the evolution of new non-equilibrium configurations investigated.

\section*{Acknowledgments}
Samuel J. Harris was supported by a UK Engineering and Physical Sciences Research Council (EPSRC) doctoral prize fellowship, grant number EP/W524414/1, at the School of Engineering Mathematics and Technology, University of Bristol.

\section*{Appendix}
\begin{appendices}
\section{Translating vortex pair WOZ algorithm}\label{app:tWOZ}
The numerical algorithm used by Wu, Overman and Zabusky (WOZ) in \cite{wu1984steady} to compute the equilibrium of two translating, counter-rotating vortex patches given some initial boundary data is here summarised. Note that since the vortex patches are symmetric about the real axis, only half of the boundary of the right vortex patch is considered i.e. $-\pi\leq\theta\leq0$ which is made up of $P+1$ boundary points. We define
\begin{equation}
    z_k-C_1 = R(\theta_k)(\cos(\theta_k)+i\sin(\theta_k))
\end{equation}
where $C_1$ is the centre of the right vortex patch. We also define $R_a = R(-\pi)$ and $R_b = R(0)$. The WOZ numerical routine is split into five steps at each iteration. 

First, the velocity on each boundary point must be found, which \cite{wu1984steady} calculate using contour dynamics techniques, see their equation (4.1). We instead use the AAALS algorithm in this work to find the velocity - note that by symmetry the velocity on each of the two boundaries is the same. Note that the AAALS algorithm requires the entire boundary data of both patches rather than just the segment of the right patch. Furthermore, if the vortex patch has changed shape following an iteration, the new shape must again be non-dimensionalised; it will have a new radius $L$ and velocity $V$ and thus the non-dimensional constant $\lambda$ must be updated.

Second, the new translational velocity $V$ is calculated
\begin{equation}\label{eq:angvelWOZ}
    V^{(n)} = \frac{1}{x_A-x_B} \sum_{k=1}^P u_{k+1/2}^{(n)}\Delta y_k^{(n)} - v_{k+1/2}^{(n)}\Delta x_k^{(n)},
\end{equation}
where $f_{k+1/2} = \frac{1}{2}[f_{k+1}+f_{k}]$ and $\Delta f_k = f_{k+1}-f_{k}$. Third, the following function is  calculated using $V$
\begin{equation}\label{eq:Ffun}
    F_{k+1/2} \equiv \frac{u_{k+1/2}\sin{\theta_{k+1}}-(v_{k+1/2}-V)\cos{\theta_{k+1}}}{u_{k+1/2}\sin{\theta_{k}}-(v_{k+1/2}-V)\cos{\theta_{k}}}.
\end{equation}

Fourth, the velocity $(u,v)$ and translational velocity $V$ are used to evaluate the kinematic boundary condition - the condition that particles on the vortex patch boundary remain on the boundary. The second order discretisation of this condition \citep{wu1984steady} can be averaged and written as 
\begin{equation}\label{eq:kinematic}
    -\frac{1}{2}F^{-1}_{k-1/2}\overline{R_{k-1}}+\overline{R_k}-\frac{1}{2}F_{k+1/2}\overline{R_{k+1}} = 0, \;\;\;\; 2\leq k\leq P,
\end{equation}
where $\overline{R_1}=R_a$, $\overline{R}_{P+1}=R_b$ and $F$ is the function given in \eqref{eq:Ffun}. The system \eqref{eq:kinematic} can be written as the product $AX=Y$, where $A$ is a tridiagonal matrix of coefficients and $Y$ is the vector $(\frac{1}{2}F^{-1}_{3/2} R_a, 0, 0, ... 0, \frac{1}{2}F_{P+1/2} R_b)$. The vector of unknowns $X=(\overline{R_2},...,\overline{R_P})$ can then be found using the MATLAB backslash operator $X = A\backslash Y$ and it follows that $\overline{R_k}\rightarrow\overline{R_k}^{(n+1)}$. The following relaxation procedure is then used
\begin{equation}
    R_k^{(n+1)}=\mu\overline{R_k}^{(n+1)}+(1-\mu)R_k^{(n)},
\end{equation}
where $\mu=0.6$ as in \cite{wu1984steady}.

Fifth, the boundary data $Z$ and segment data $Z_s$ are updated accordingly. Then, the iteration continues until the following error criterion is satisfied
\begin{equation}\label{eq:WOZtoly}
    \sum_{k=1}^{P+1}|R_k^{(n+1)}-R_k^{(n)}|<\epsilon,
\end{equation}
for some tolerance $\epsilon=5\times10^{-7}$ or until some maximum number of iterations $I_{max}$ is achieved; such an $I_{max}$ is implemented to avoid exceptionally long runtimes. In this work, the value $I_{max}=500$ was used, yet all results presented in this work converged to an equilibrium in fewer iterations.

\section{Rotating multipolar vortices WOZ algorithm}\label{app:mrWOZ}
Following \cite{dritschel1985stability} and \cite{xue2017new}, see appendix C of the latter, we modify the WOZ algorithm given in \cite{wu1984steady} for calculating multipolar vortex patch equilibria. Note that the maximal central patch radius $a$, the satellite inner and outer distances $b_1$ and $b_2$ (see figure \ref{fig:tr_vortex_patch}, right) and the vorticity of the central and satellite patches are fixed, with the positions of the other boundary points free to vary. In addition, only a $1/N$ segment of the central patch and one half of one of the satellite patches are required by symmetry.

The iterative scheme is very similar to that given in appendix \ref{app:tWOZ} with some notable changes. We define
\begin{equation}
    z_k-C_n = R(\theta_k)(\cos(\theta_k)+i\sin(\theta_k))
\end{equation}
where $C_n=x_n+iy_n$ is the centre of vortex patch $n$. Step 2 of the algorithm requires the calculation of the angular velocity $\Omega$ using the following formula
\begin{equation}\label{eq:Omformula}
    \Omega^{(n)} = \frac{2}{b_1^2-b_2^2}\sum_{k=1}^{P+1} u_{k+1/2}^{(n)}\Delta y_k^{(n)} - v_{k+1/2}^{(n)}\Delta x_k^{(n)}.
\end{equation}
The quantity $F_{n,k+1/2}$ in step 3 must then be calculated for each patch (meaning the central patch and one satellite patch by symmetry) as follows
\begin{equation}
    F_{n, k+1/2} \equiv \frac{u_{k+1/2}\sin{\theta_{k+1}}-v_{k+1/2}\cos{\theta_{k+1}}+(\Omega/2)R_{k+1}+\Omega x_n\cos(\theta_{k+1})+\Omega y_n\sin(\theta_{k+1})}{u_{k+1/2}\sin{\theta_{k}}-v_{k+1/2}\cos{\theta_{k}}+(\Omega/2)R_{k}+\Omega x_n\cos\theta_{k}+\Omega y_n\sin\theta_{k}}.
\end{equation}
The remaining steps of the algorithm in appendix \ref{app:tWOZ} are unchanged, except that a tolerance of $\epsilon=5\times10^{-5}$ is used for the rotating multipolar case. Note that this algorithm is for the case of vortex patches with same-signed vorticity of equal magnitude. For patches with unequal vorticity, \eqref{eq:Omformula} requires some slight modifications - see appendix C of \cite{xue2017new}.

\section{Analytical solutions of the two- and three-contour Rankine vortices}\label{app:analy}
We here state the analytical solution for the velocity field of the two- and three-contour Rankine vortices introduced in section \ref{sec:nested}. These solutions are lifted from \cite{morel1994multipolar}, noting that the labelling of regions in the current work is different to the labelling used in \cite{morel1994multipolar}. The velocity field (in complex notation) of the three-contour Rankine vortex is
\begin{equation}
    u+iv = 
    \begin{cases}
      \Gamma_t/(2r), & |z|\geq A_2 \text{ (region 0, exterior flow);}\\
      \gamma B^2/(2r) + r/2 - A_1^2/(2r), & A_2\geq|z|\geq A_1 \text{ (region 1);}\\
      \gamma B^2/(2r),  & A_1\geq|z|\geq B \text{ (region 2);}\\
      \gamma r/2, & B\geq|z| \text{ (region 3),}
    \end{cases}   
\end{equation}
where $\Gamma_t=\gamma B^2 +A_2^2-A_1^2$ and $\gamma<0$ is the vorticity of the central patch. Note that the equivalent result for the two-contour Rankine vortex can be obtained by considering the limit $A_1\rightarrow B$.

\end{appendices}

\bibliographystyle{unsrt}  
\bibliography{ABbib}

\end{document}